# Four-parameter Mittag-Leffler functions and their associated coherent states

**Dušan POPOV**

University Politehnica of Timisoara, Romania
Department of Physical Foundations of Engineering
Bd. V. Parvan No. 2, 300223 Timisoara, Romania
E-mail: dusan_popov@yahoo.co.uk
**ORCID:** https://orcid.org/0000-0003-3631-3247

**Abstract**

We defined and used a pair of Hermitian annihilation and creation operators which generate the generalized coherent states, defined in the Barut-Girardello manner, whose normalization function is just the four-parameter generalized Mittag-Leffler function. We examined the characteristic properties for these pure, as well as mixed (thermal) coherent states. All calculations are made using the rules of the technique of diagonal ordering of operators. Finally, the integral counterpart of the Mittag-Leffler coherent states are examined which is connected with *nu*-function. The paper is an example of the application of a mathematical entity (Mittag-Leffler function) in quantum mechanics (coherent states formalism).

**1. Introduction and preliminaries**

The Mittag-Leffler function was introduced in 1903 [1] as an entire one parametric complex special function which depends on some complex parameter $\alpha$. Later, the definition was extended, adding one, two, three or more complex parameters $\beta$, $\gamma$, $k$ …. The history of this function can be traced in [2], [3]:

$$E_\alpha(z) = \sum_{n=0}^{\infty} \frac{z^n}{\Gamma(1+\alpha n)} \ , \quad E_{\alpha,\beta}(z) = \sum_{n=0}^{\infty} \frac{z^n}{\Gamma(\beta+\alpha n)} \ , \quad E_{\alpha,\beta}^{\gamma}(z) = \sum_{n=0}^{\infty} \frac{(\gamma)_n}{\Gamma(\beta+\alpha n)} \frac{z^n}{n!}, \quad (1.1)$$

with $z, \alpha, \beta, \gamma \in C$, $Re(\alpha) > 0$.

Moreover, a superior step of generalization was recently considered by Srivastava and Tomovski [4], the form that we will deal with in this paper. The only difference that we have introduced, compared to the Srivastava and Tomovski notation, is the exchange of the places of the lower indices $\alpha$ and $\beta$, because of the roles they play in the definition of the Mittag-Leffler function, index $\alpha$ being similar to index $k$. So we will use $E^{\gamma,k}_{\beta,\alpha}(z)$ instead of $E^{\gamma,k}_{\alpha,\beta}(z)$:

$$E^{\gamma,k}_{\beta,\alpha}(z) = \sum_{n=0}^{\infty} \frac{(\gamma)_{n,k}}{\Gamma_\alpha(\beta+\alpha n)} \frac{z^n}{n!}, \quad z,\alpha,\beta,\gamma \in C, \; \mathcal{R}e(\alpha)>0, \; \mathcal{R}e(k)>0 \tag{1.2}$$

In recent years, other generalizations of the Mittag-Leffler function, characterized by several indices, have been developed (see, for example, [5] and the references therein).

In the present paper we will limit ourselves to the Srivastava and Tomovski version, which we will briefly call hereafter the *generalized Mittag-Leffler function* (GM-L). In order to avoid possible confusion, compared to the articles in the specialized literature, we inserted, in the writing of the GM-L function, the $\alpha$ index signifying the rate of change of the argument of the k-Gamma function.

In the above definition appear the generalized Pochhammer symbols (or Pochhammer $k$-symbols), defined as [6]

$$(\gamma)_{n,k} = \gamma(\gamma+k)(\gamma+2k)...(\gamma+(n-1)k) = \frac{\Gamma_k(\gamma+nk)}{\Gamma(\gamma)} \tag{1.3}$$

where $\Gamma_k(\gamma+nk)$ is the generalized $k$ – gamma function, defined by the integral, or $\mathcal{R}e(k)>0$:

$$\Gamma_k(x) = \int_0^\infty dt\, e^{-\frac{t^k}{k}} t^{x-1}, \qquad \Gamma_k(x) = k^{\frac{x}{k}-1}\Gamma\left(\frac{x}{k}\right) \tag{1.4}$$

The connection between the generalized Pochhammer $k$ – symbol and usual Pochhammer symbol is

$$(x)_{n,k} = k^n \left(\frac{x}{k}\right)_n = k^n \frac{\Gamma\left(\frac{x}{k}+n\right)}{\Gamma\left(\frac{x}{k}\right)}, \quad (x)_{0,k} = \left(\frac{x}{k}\right)_0 = 1, \quad (x)_{n,0} = x^n \tag{1.5}$$

As well as, the generalized gamma function $\Gamma_\alpha(\beta+\alpha n)$ is connected with usual Gamma function as

$$\Gamma_\alpha(\beta+\alpha n) = \alpha^n \left(\frac{\beta}{\alpha}\right)_n \Gamma(\beta) = \Gamma(\beta)(\beta)_{n,\alpha} \tag{1.6}$$

So, the generalized Mittag-Leffler function becomes

$$E^{\gamma,k}_{\beta,\alpha}(z) = \frac{1}{\Gamma(\beta)} \sum_{n=0}^{\infty} \frac{(\gamma)_{n,k}}{(\beta)_{n,\alpha}} \frac{z^n}{n!} \tag{1.7}$$

Shortly, we will call this function the *generalized Mittag-Leffler function* (GM-L), and it can be linked with a generalized hypergeometric function (GHG), as:

$$E^{\gamma,k}_{\beta,\alpha}(z) = \frac{1}{\Gamma(\beta)} \, _1F_1\left(\frac{\gamma}{k}; \frac{\beta}{\alpha}; \frac{k}{\alpha}z\right) \tag{1.8}$$

The purpose of this paper is to build such a type of coherent states (CSs) that has the generalized Mittag-Leffler function as its normalization function. Hereby we will highlight a new application of the generalized Mittag-Leffler function, in addition to those indicated in the literature [2], [7], [8]. But as the normalization function of the generalized CSs is a generalized hypergeometric function, by default we will capitalize on the connection between the generalized Mittag-Leffler function and a certain generalized hypergeometric function, materialized by the relation above.

Moreover, the most general structure of a certain kind of CSs $|z>$ as the development in the Fock-vectors orthogonal basis $\{|n>, n=0,1,2,...,\infty\}$ is

$$|z> = \frac{1}{\sqrt{{}_pF_q(\{a_i\}_1^p; \{b_j\}_1^q; |z|^2)}} \sum_{n=0}^{\infty} \frac{z^n}{\sqrt{\rho_{p,q}(n)}} |n> \qquad (1.9)$$

where the complex variable $z=|z|\exp(i\varphi)$ labels the CSs, $0 \leq \varphi \leq 2\pi$, and where the positive quantities $\rho_{p,q}(n)$, called the structure constants, determine the mathematical structure of CSs.

The normalization function of the most general CSs is a hypergeometric generalized function (HGF) ${}_pF_q(\{a_i\}_1^p; \{b_j\}_1^q; |z|^2)$, where $\{x_i\}_1^m \equiv x_1, x_2, x_3,..., x_m$ and so on.

These kinds of states were firstly introduced by Appl and Schiller [9] and applied to the thermal states of the pseudoharmonic oscillator in one of our previous papers [10].
Any CSs exists (makes physical sense) only if their normalization function is an analytical function of real variable $|z|^2$. In addition, any CSs $|z>$ must accomplish some conditions [11]: continuity in complex label, normalization, non-orthogonality, unity operator resolution with unique positive weight function $h(|z|^2)$ of the integration measure $d\mu(z)$.

The very important relation in the CSs formalism is the *resolution of the unity operator*, or the *completeness relation*, i.e. the integral

$$\int d\mu(z) |z><z| = 1 \ . \qquad (1.10)$$

where the integration measure $d\mu(z) = \frac{d^2z}{\pi} h(|z|^2) = \frac{d\varphi}{2\pi} d(|z|^2) h(|z|^2)$ must have positive, unique and no oscillatory weight function $h(|z|^2)$.

The generalized coherent states are generated with the help of a pair of operators of creation $\mathcal{A}_+$ and annihilation $\mathcal{A}_+$ that act on the Fock vectors $|n>$ in the space attached to the studied quantum system. They obey the rules of a normal ordering technique - diagonal ordering operation technique (DOOT), which consist of the following [12]: *a)* Inside the symbol # # the order of the operators $\mathcal{A}_-$ and $\mathcal{A}_+$ can be permuted like commutable operators, so that finally will result a normal ordering: $\mathcal{A}_+$ on the left, and $\mathcal{A}_-$ on the right; *b)* A symbol # # inside another symbol # # can be deleted; *c)* A normally ordered product of operators can be integrated or differentiated, with respect to *c*-numbers, according to the usual rules. In addition, the operators $\mathcal{A}_-$ and $\mathcal{A}_+$ are considered as being simple *c*-numbers and can be taken out from the symbol # #; *d)* The projector $|0><0|$ of the normalized vacuum state $|0>$, in the frame of DOOT, is the reciprocal function $(1/f(x))$ of the normalization function of CSs, in normal order, i. e. the reciprocal function of normal ordered generalized hypergeometric function:

$$|0><0| = \# \frac{1}{{}_pF_q(\{a_i\}_1^p; \{b_j\}_1^q; A_+A_-)} \# \qquad (1.11)$$

4## 2. Coherent states of the generalized Mittag-Leffler functions

Let's point out that, for the first time, the CSs for the ordinary one-parametrical and two-parametrical Mittag-Leffler functions were examined by Sixdeniers, Penson and Solomon [13].

Now, let us we choose two Hermitian generating operators, lowering $\mathcal{A}_{-(\beta,\alpha)}^{(\gamma,k)} \equiv \mathcal{A}_-$ and raising $\mathcal{A}_{+(\beta,\alpha)}^{(\gamma,k)} \equiv \mathcal{A}_+$, where $(\mathcal{A}_-)^+ = \mathcal{A}_+$. In order not to load the writing of the formulas, we will continue to use the short version of the operators, $\mathcal{A}_-$ and $\mathcal{A}_+$. The pair of operators was chosen in such a way that their action on the Fock vectors is of the following form:

$$\mathcal{A}_-|n> = \sqrt{e(n)}|n-1>, \quad \mathcal{A}_+|n> = \sqrt{e(n+1)}|n+1>, \quad e(n) \equiv n\frac{(\beta+\alpha(n-1))}{\gamma+k(n-1)} \qquad (2.1)$$

which, as will be seen below, will generate the coherent states of the generalized Mittag-Leffler functions (GML-CSs).

By successively applying these operators on the vacuum states $|0>$ and $<0|$ we obtain

$$\begin{Bmatrix} |n> \\ <n| \end{Bmatrix} = \sqrt{\frac{(\gamma)_{n,k}}{(\beta)_{n,\alpha}\, n!}} \begin{Bmatrix} (\mathcal{A}_+)^n |0> \\ <0|(\mathcal{A}_-)^n \end{Bmatrix} = \sqrt{\Gamma(\beta)}\sqrt{\frac{(\gamma)_{n,k}}{\Gamma_\alpha(\beta+\alpha n)}} \begin{Bmatrix} (\mathcal{A}_+)^n |0> \\ <0|(\mathcal{A}_-)^n \end{Bmatrix} \qquad (2.2)$$

Using the completeness relation of the Fock-vectors $\sum_{n=0}^{\infty}|n><n|=1$ and the rules of the DOOT technique, we obtain the vacuum projector:

$$\sum_{n=0}^{\infty}|n><n| = \sum_{n=0}^{\infty}\frac{(\gamma)_{n,k}}{(\beta)_{n,\alpha}\, n!} \#(\mathcal{A}_+)^n|0><0|(\mathcal{A}_-)^n\# =$$

$$= |0><0|\sum_{n=0}^{\infty}\frac{(\gamma)_{n,k}}{(\beta)_{n,\alpha}}\frac{\#(\mathcal{A}_+\mathcal{A}_-)^n\#}{n!} = |0><0|\Gamma(\beta)\# E_{\alpha,\beta}^{\gamma,k}(\mathcal{A}_+\mathcal{A}_-)\# = 1 \qquad (2.3)$$

From here it follows that the vacuum projector is

$$|0><0| = \frac{1}{\Gamma(\beta)\# E_{\beta,\alpha}^{\gamma,k}(\mathcal{A}_+\mathcal{A}_-)\#} = \frac{1}{\#_1 F_1\left(\frac{\gamma}{k};\frac{\beta}{\alpha};\frac{k}{\alpha}\mathcal{A}_+\mathcal{A}_-\right)\#} \qquad (2.4)$$

in accordance with DOOT rule *d)*. This means that the CSs normalization function for GML-CSs is just the four-parametrical Mittang-Leffler function. Let's check this assertion by actually building CSs of the GM-L functions.

First, we will use the Barut Girardello definition of coherent states [14]:

$$\mathcal{A}_-|z> = z|z>, \quad <z|\mathcal{A}_+ = z^*<z| \qquad (2.5)$$

Using this definition, as well as applying the norming condition $<z|z>=1$, the development of CSs according to the Fock vectors $|z> = \sum_n c_n(z)|n>$ leads to the expression

$$|z> = \frac{1}{\sqrt{E_{\beta,\alpha}^{\gamma,k}(|z|^2)}} \sum_{n=0}^{\infty} \sqrt{\frac{(\gamma)_{n,k}}{\Gamma_\alpha(\beta+\alpha n)}} \frac{z^n}{\sqrt{n!}} |n> \qquad (2.6)$$

Using Eq. (2.2), the Mittag-Leffler CSs can be written also as

$$|z> = \sqrt{\Gamma(\beta)} \frac{1}{\sqrt{E_{\beta,\alpha}^{\gamma,k}(|z|^2)}} E_{\beta,\alpha}^{\gamma,k}(z\mathcal{A}_+)|0> \qquad (2.7)$$

It can be seen that the generalized Mittag-Leffler function plays the role of CSs normalization function.

Let's verify that this expression of CSs satisfies Klauder's conditions [13]: continuity in complex label, normalization, non-orthogonality, unity operator resolution with unique positive weight function $h(|z|^2)$ of the integration measure $d\mu(z)$.

The vectors $|z>$ are the strong continuous functions of label $z \neq 0$, i.e. for every convergent label such that $z' \to z$ it follows that $\||z'> - |z>\| \to 0$.

From the expression of the scalar product of two CSs we can see the properties of normalization and non-orthogonality:

$$<z|z'> = \frac{E_{\beta,\alpha}^{\gamma,k}(z^*z')}{\sqrt{E_{\beta,\alpha}^{\gamma,k}(|z|^2) E_{\beta,\alpha}^{\gamma,k}(|z'|^2)}} = \begin{cases} 1, & \text{if } z' = z \quad \text{normalization to unity} \\ \neq 0, & \text{if } z' \neq z \quad \text{non-orthogonality} \end{cases} \qquad (2.8)$$

To show the validity of the unity operator decomposition relation (1.10), we will try to find the expression of the integration measure $h(|z|^2)$. After the corresponding substitutions, the angular integration

$$\int_0^{2\pi} \frac{d\varphi}{2\pi} (z^*)^n z^{n'} = (|z|^2)^n \delta_{nn'} \qquad (2.9)$$

and the shifting of the index $n = s-1$, we will have to solve the following integral equation (problem of moments) [14]:

$$\int_0^\infty d(|z|^2) \frac{h(|z|^2)}{E_{\beta,\alpha}^{\gamma,k}(|z|^2)} (|z|^2)^{s-1} = \frac{k}{\alpha} \frac{\Gamma\left(\frac{\gamma}{k}\right)}{\Gamma\left(\frac{\beta}{\alpha}\right)} \Gamma(\beta) \frac{1}{\left(\frac{k}{\alpha}\right)^s} \frac{\Gamma(s)\Gamma\left(\frac{\beta}{\alpha}-1+s\right)}{\Gamma\left(\frac{\gamma}{k}-1+s\right)} \qquad (2.10)$$

The solution of this equation being expressed through Meijer's G-function [14]

$$\frac{h(|z|^2)}{E_{\beta,\alpha}^{\gamma,k}(|z|^2)} = \frac{k}{\alpha} \frac{\Gamma\left(\frac{\gamma}{k}\right)}{\Gamma\left(\frac{\beta}{\alpha}\right)} \Gamma(\beta) G_{1,2}^{2,0}\left(\frac{k}{\alpha}|z|^2 \left| \begin{array}{c} /\,; \quad \frac{\gamma}{k}-1 \\ 0, \frac{\beta}{\alpha}-1; \quad / \end{array} \right.\right) \qquad (2.11)$$

the final expression for the integration measure is

$$d\mu(z) = \frac{k}{\alpha} \frac{\Gamma\left(\frac{\gamma}{k}\right)}{\Gamma\left(\frac{\beta}{\alpha}\right)} \Gamma(\beta) \frac{d\varphi}{2\pi} d(|z|^2) E_{\beta,\alpha}^{\gamma,k}(|z|^2) G_{1,2}^{2,0}\left(\frac{k}{\alpha}|z|^2 \left|\begin{array}{c} /\ ;\ \frac{\gamma}{k}-1 \\ 0,\frac{\beta}{\alpha}-1;\ / \end{array}\right.\right) \qquad (2.12)$$

In this way, we showed that Mittag-Leffler CSs fulfill all Klauder's conditions.

*Additional note 1*: For the particular case, when $\alpha = \beta = \gamma = k = 1$, we will have $E_{1,1}^{1,1}(|z|^2) = \exp(|z|^2)$ and $G_{0,1}^{1,0}(|z|^2|0) = \exp(-|z|^2)$ so that the integration measure of the one-dimensional harmonic oscillator (HO-1D) is reached: $d\mu(z) = \frac{d\varphi}{2\pi} d(|z|^2) = \frac{d^2z}{\pi}$, which proves the correctness of the above calculation.

Let's also point out that during the above calculation we used the classical and generalized integral of a single Meijer G-function, which will be useful in the continuation of the paper:

$$\int_0^\infty d(|z|^2)(|z|^2)^{s-1} G_{1,2}^{2,0}\left(\frac{k}{\alpha}|z|^2 \left|\begin{array}{c} /\ ;\ \frac{\gamma}{k}-1 \\ 0,\frac{\beta}{\alpha}-1;\ / \end{array}\right.\right) = \frac{1}{\left(\frac{k}{\alpha}\right)^s} \frac{\Gamma(s)\Gamma\left(\frac{\beta}{\alpha}-1+s\right)}{\Gamma\left(\frac{\gamma}{k}-1+s\right)} \qquad (2.13)$$

Generally, with the help of DOOT rules, the projector onto the one CS $|z>$ is

$$|z><z| = \frac{1}{E_{\beta,\alpha}^{\gamma,k}(|z|^2)} \# \frac{E_{\beta,\alpha}^{\gamma,k}(\mathcal{A}_+ z) E_{\beta,\alpha}^{\gamma,k}(\mathcal{A}_- z^*)}{E_{\beta,\alpha}^{\gamma,k}(\mathcal{A}_+ \mathcal{A}_-)} \# \qquad (2.14)$$

Using the limit

$$\lim_{z \to 0} E_{\beta,\alpha}^{\gamma,k}(z) = \frac{1}{\Gamma(\beta)} \qquad (2.15)$$

as well as the completeness relation, we will obtain, on the one hand, the correct expression of the vacuum projector, Eq. (2.14), and on the other hand, the following two integrals:

$$\int_0^{2\pi} \frac{d\varphi}{2\pi} \# E_{\beta,\alpha}^{\gamma,k}(\mathcal{A}_+ z) E_{\beta,\alpha}^{\gamma,k}(\mathcal{A}_- z^*) \# = \sum_{n=0}^\infty \left[\frac{(\gamma)_{n,k}}{\Gamma_\alpha(\beta+\alpha n)} \frac{1}{n!}\right]^2 \#(\mathcal{A}_+ \mathcal{A}_-)^n \#(|z|^2) \qquad (2.16)$$

$$\int \frac{d^2z}{\pi} G_{1,2}^{2,0}\left(\frac{k}{\alpha}|z|^2 \left|\begin{array}{c} /\ ;\ \frac{\gamma}{k}-1 \\ 0,\frac{\beta}{\alpha}-1;\ / \end{array}\right.\right) \# E_{\beta,\alpha}^{\gamma,k}(\mathcal{A}_+ z) E_{\beta,\alpha}^{\gamma,k}(\mathcal{A}_- z^*) \# =$$
$$= \frac{\alpha}{k} \frac{\Gamma\left(\frac{\gamma}{k}\right)}{\Gamma\left(\frac{\beta}{\alpha}\right)} \frac{1}{\Gamma(\beta)} \# E_{\beta,\alpha}^{\gamma,k}(\mathcal{A}_+ \mathcal{A}_-) \# \qquad (2.17)$$

The Laplace type integral, or the Laplace transform of the generalized Mittag-Leffler function is

$$\int_0^\infty \frac{d^2 z}{\pi} e^{-s|z|^2} E_{\beta,\alpha}^{\gamma,k}(|z|^2) = \frac{1}{s} \frac{1}{\Gamma\left(\frac{\gamma}{k}\right)} \frac{1}{\Gamma(\beta)} \,_2F_1\left(1, \frac{\gamma}{k}\,;\, \frac{\beta}{\alpha}\,;\, \frac{k}{\alpha}\frac{1}{s}\right) \quad (2.18)$$

***Additional note 2:*** For $\alpha = \beta = \gamma = k = 1$, we have $E_{1,1}^{1,1}(|z|^2) = \exp(|z|^2)$ and the Laplace transform becomes an integral of an exponential having the value $(1-s)^{-1}$.

Taking into account the definition of coherent states in the Barut-Girardello manner (BG-CSs), for a function that depends on the ordered product of $\#f(\mathcal{A}_+\mathcal{A}_-)\#$ operators, following the DOOT rules, it can be shown that its expected value in the coherent state $|z>$ is:

$$<z|\#f(\mathcal{A}_+\mathcal{A}_-)\#|z> = f(|z|^2) \quad (2.19)$$

***Additional note 3:*** Coherent states can also be defined in the *Klauder-Perelomov manner*, that is, as the result of the action of the displacement operator on the vacuum state [15]:

$$|z> = \frac{1}{\sqrt{N(|z|^2)}} \exp(z\mathcal{A}_+)|0> = \frac{1}{\sqrt{_qF_p\left(\{b_j\}_1^q\,;\,\{a_i\}_1^p\,;\,|z|^2\right)}} \sum_{n=0}^\infty \frac{z^n}{\sqrt{\rho_{q,p}(n)}}|n> \quad (2.20)$$

From an algebraic point of view, CSs in the Klauder-Perelomov manner (CSs) have the same structure as CSs in the Barut-Girardello manner (BG-CSs), with the fundamental difference that there is an interchange of the places of the indices $p$ and $q$, as well as of the sets of constants $\{a_i\}_1^p$ and $\{b_j\}_1^q$. This is an expression of the dualism of the two types of CSs. Therefore, the expression obtained above for BG-CSs are similar for KP-CSs, so we will not repeat them.

### 3. Generalized Mittag-Leffler functions involved in thermal states

It is well known from quantum mechanics books that a system in thermal equilibrium at temperature $T = (k_B \beta_B)^{-1}$ with the ambient medium is characterized by the canonical density operator, where $B$ = Boltzmann :

$$\rho = \frac{1}{Z(\beta_B)} \sum_{n=0}^\infty \exp(-\beta_B E_n)|n><n| \quad (3.1)$$

Using Eqs. (2.3) and (2.4) and the DOOT rules, the density operator $\rho$ becomes a function of the operator product $\mathcal{A}_+\mathcal{A}_-$:

$$\rho = \frac{1}{Z(\beta_B)} \frac{1}{\#E_{\beta,\alpha}^{\gamma,k}(\mathcal{A}_+\mathcal{A}_-)\#} \sum_{n=0}^\infty \exp(-\beta_B E_n) \frac{(\gamma)_{n,k}}{\Gamma_\alpha(\beta+\alpha n)} \frac{\#(\mathcal{A}_+\mathcal{A}_-)^n\#}{n!} \quad (3.2)$$

Let's turn our attention to the quantity defined as follows, Eq. (2.1):

$$e(n) \equiv n \frac{\beta + \alpha(n-1)}{\gamma + k(n-1)} \quad (3.3)$$

For relatively simple quantum systems, the energy spectrum is either linear or quadratic, in relation to the main quantum number $n$.

If $\alpha = 0$ and $k = 0$, we have $e(n) = \dfrac{\beta}{\gamma} n$, which corresponds to a *linear spectrum* (characteristic for the HO-1D linear harmonic oscillator or the PHO pseudoharmonic oscillator).

If $\alpha \neq 0$ and $k = 0$, we have $e(n) = \dfrac{\beta}{\gamma}(1-\alpha)n + \dfrac{\beta}{\gamma}\alpha n^2$, which corresponds to a *quadratic spectrum* (characteristic for the nonlinear or anharmonic harmonic oscillators, like Morse oscillator MO or some Pöschl-Teller-type potentials and so on.

For *linear spectra* the density operator becomes

$$\rho_{\text{lin}} = \frac{1}{Z(\beta_B)} \frac{1}{\Gamma(\beta_B)} \frac{1}{\# E_{\beta,\alpha}^{\gamma,k}(\mathcal{A}_+\mathcal{A}_-)\#} \sum_{n=0}^{\infty} \frac{(\gamma)_{n,k}}{(\beta)_{n,\alpha}} \frac{\# \left(e^{-\beta_B \frac{\beta}{\gamma}} \mathcal{A}_+\mathcal{A}_-\right)^n \#}{n!} = \tag{3.4}$$

$$= \frac{1}{Z(\beta_B)} \frac{1}{\# E_{\beta,\alpha}^{\gamma,k}(\mathcal{A}_+\mathcal{A}_-)\#} \# E_{\beta,\alpha}^{\gamma,k}\left(e^{-\beta_B \frac{\beta}{\gamma}} \mathcal{A}_+\mathcal{A}_-\right) \#$$

Generally, the Husimi's function (or Husimi's distribution) is defined as the diagonal elements of the density operator, in the CSs representation.

In our case, for the Mittag-Leffler CSs, the Husimi's distribution becomes

$$Q_{ML}(|z|^2) \equiv \langle z | \rho_{\text{lin}} | z \rangle = \frac{1}{Z(\beta_B)} \frac{1}{E_{\beta,\alpha}^{\gamma,k}(|z|^2)} E_{\beta,\alpha}^{\gamma,k}\left(e^{-\beta_B \frac{\beta}{\gamma}} |z|^2\right) \tag{3.5}$$

It can be observed that, compared to the expression above, $\mathcal{A}_+\mathcal{A}_-$ has been replaced with $|z|^2$, as a consequence of the definition of CSs in the sense of Barut and Girardello.

On the other hand, the $\rho_{\text{lin}}$ density operator can also be written in diagonal form:

$$\rho_{\text{lin}} = \int d\mu(z) |z\rangle P_{\text{lin}}(|z|^2) \langle z| \tag{3.6}$$

$$\rho_{\text{lin}} = \frac{1}{E_{\beta,\alpha}^{\gamma,k}(\mathcal{A}_+\mathcal{A}_-)} \int d\mu(z) \, P_{\text{lin}}(|z|^2) \# \frac{E_{\beta,\alpha}^{\gamma,k}(\mathcal{A}_+ z) E_{\beta,\alpha}^{\gamma,k}(\mathcal{A}_- z^*)}{E_{\beta,\alpha}^{\gamma,k}(|z|^2)} \# \tag{3.7}$$

and it is necessary to find the quasi-distribution function $P_{\text{lin}}(|z|^2)$.

Equating this equation with the second row of Eq. (3.4), substituting the expressions of the integration measure and CSs, after performing the angular integral, we have to solve the following equality

$$\frac{1}{Z(\beta_B)} \# E_{\beta,\alpha}^{\gamma,k}\left(e^{-\beta_B \frac{\beta}{\gamma}} \mathcal{A}_+ \mathcal{A}_-\right) \# = \sum_{n=0}^{\infty} \left[ \frac{(\gamma)_{n,k}}{\Gamma_\alpha(\beta + \alpha n)} \frac{\#\left(e^{-\beta_B \frac{\beta}{\gamma}} \mathcal{A}_+ \mathcal{A}_-\right)^n \#}{n!} \right] \times$$

$$\times \frac{1}{\left(\frac{\alpha}{k}\right)^n} \frac{1}{\left(e^{+\beta_B \frac{\beta}{\gamma}}\right)^n} \frac{1}{\Gamma(n+1)} \frac{\Gamma\left(\frac{\gamma}{k}+n\right)}{\Gamma\left(\frac{\beta}{\alpha}+n\right)} \int_0^\infty d(|z|^2)(|z|^2)^n \, G_{1,2}^{2,0}\left(\frac{k}{\alpha}|z|^2 \Big| \ldots\right) P_{\text{lin}}(|z|^2)$$

(3.8)

Here it appears an integral equation similar with Eq. (2.13), but in what unknown function it is now $G_{1,2}^{2,0}\left(\frac{k}{\alpha}|z|^2 \Big| \ldots\right) P_{\text{lin}}(|z|^2)$.

Proceeding similarly as for the case of finding the integration measure, the final solution for the quasi-distribution function will be

$$P_{\text{lin}}(|z|^2) = \frac{1}{Z(\beta_B)} e^{\beta_B \frac{\beta}{\gamma}} \frac{G_{1,2}^{2,0}\left(\frac{k}{\alpha} e^{\beta_B \frac{\beta}{\gamma}}|z|^2 \left| \begin{array}{c} /\,; \quad \frac{\gamma}{k}-1 \\ 0\,,\, \frac{\beta}{\alpha}-1;\, / \end{array} \right.\right)}{G_{1,2}^{2,0}\left(\frac{k}{\alpha}|z|^2 \left| \begin{array}{c} /\,; \quad \frac{\gamma}{k}-1 \\ 0\,,\, \frac{\beta}{\alpha}-1;\, / \end{array} \right.\right)}$$

(3.9)

For *quadratic* energy spectra, if the following constants are $\alpha \neq 0$ and $k = 0$, we have $E(n) \equiv e(n) = \frac{\beta}{\gamma}(1-\alpha)n + \frac{\beta}{\gamma}\alpha n^2 \equiv An + Bn^2$ (where we used that $\hbar\omega = 1$), and a specific *ansatz* is applied that we used for the first time in paper [16]. It consists in the power series development of the quadratic part of the energy exponential and the application of the above calculation for the linear part:

$$e^{-\beta_B E(n)} = e^{-\beta_B B n^2} \left(e^{-\beta_B A}\right)^n = \sum_{j=0}^{\infty} \frac{(-\beta_B B)^j}{j!} n^{2j} \left(e^{-\beta_B A}\right)^n =$$

$$= \sum_{j=0}^{\infty} \frac{(-\beta_B B)^j}{j!} \left(\frac{\partial}{\partial \beta_B A}\right)^{2j} \left(e^{-\beta_B A}\right)^n = \exp\left[-\frac{1}{\beta_B} B \left(\frac{\partial}{\partial A}\right)^2\right] \left(e^{-\beta_B A}\right)^n$$

(3.10)

In this case, the density operator becomes

$$\rho = \frac{1}{Z(\beta_B)} \frac{1}{\# E_{\beta,\alpha}^{\gamma,k}(\mathcal{A}_+ \mathcal{A}_-) \#} \exp\left[-\frac{1}{\beta_B} B \left(\frac{\partial}{\partial A}\right)^2\right] \# E_{\beta,\alpha}^{\gamma,k}\left(e^{-\beta_B A} \mathcal{A}_+ \mathcal{A}_-\right)^n \#$$

(3.11)

and the partition function is

$$Z(\beta_B) = \exp\left[-\frac{1}{\beta_B}B\left(\frac{\partial}{\partial A}\right)^2\right]\sum_{n=0}^{\infty}\left(e^{-\beta_B A}\right)^n = \exp\left[-\frac{1}{\beta_B}B\left(\frac{\partial}{\partial A}\right)^2\right]\frac{1}{1-e^{-\beta_B A}} \qquad (3.12)$$

In practical calculations, only the first few terms are retained from the infinite sum, depending on the desire for accuracy of the calculations.

## 4. Generalized Mittag-Leffler functions for continuous spectra

Let us now deal with the case of continuous spectrum systems, in which, at the limit, the difference between the "energy levels" is practically zero $\Delta E \equiv E_{n+1} - E_n = \varepsilon <<$. Consequently, we can consider $E_n = n\varepsilon$, with $\varepsilon = 1$.

Also, the coherent states for the continuous spectra were defined in [17] (see, also [18] and references therein) and [19].

The transition from discrete CSs (discontinuous) to continuous CSs is made if we adopt the following limit $d - c$. In paper [19] we studied the transition from the *discontinuous spectrum* (*d*) to the *continuous spectrum* (*c*) of an certain quantum system. We found that if a certain limit, called *the discrete – continuous limit* $d \to c$, is applied, a quantity that characterizes a system with a discontinuous spectrum will pass, at this limit, into the corresponding quantity connected with the continuous spectrum. There are some systems that have both a discontinuous and a continuous spectrum (for example, the diatomic molecule, whose inter nuclear potential is a Morse-type potential).

In the next we will adopt a following *discrete – continuous limit* $d \to c$ limit:

$$\tilde{X}_c(E) = \lim_{\substack{n \to E \\ n_{max} \to \infty \\ \sum_{n=0}^{\infty} \to \int_0^{\infty} dE}} X_d(n, n_{max}) \equiv \lim_{d \to c} X_d(n, n_{max}) \ , \qquad \sum_{n=0}^{n_{max}} X_d(n, n_{max}) \to \int_0^{\infty} dE\, \tilde{X}_c(E) \qquad (4.1)$$

So, all observables $\tilde{X}_c$ what characterizes the system with continuous spectrum will be obtained as a limiting case of the corresponding observables $X_d$ of the discrete spectrum, through three operations: *a)* replacing $n \to E$, by the dimensionless energy $E$; *b)* the extension $n_{max} \to \infty$; *c)* simultaneously, the sum with respect to $n$ must be replaced by the integral with respect to $E$. For this reason we will call this the *generalized discrete – continuous limit* $d \to c$ *limit* (Gd-cL).

In order to distinguish the observables related to the discrete spectrum $X_d$, from those of the continuous spectrum $\tilde{X}_c$, we will adopt the following notation $\tilde{X}_c \equiv \tilde{X}$ (that is, we will use the sign "tilda", i.e. the "horizontal" integral sign).

Let us we consider $E$ a dimensionless energy eigenvalue of the dimensionless Hamiltonian $\mathcal{H}$ *with a non-degenerate continuous spectrum,* and eigenstates $|E>$, $\hbar\omega = 1$, (with $0 \le E \le \infty$), i.e. $\mathcal{H}|E> = E|E>$ which are formal delta-function normalized, i.e., with $<E|E'> = \delta(E - E')$.

The closure or completeness relation for continuous spectrum is

$$\int_0^\infty dE\,|E\rangle\langle E|=1\quad,\quad \int_0^\infty dE\,\langle E'|E\rangle\langle E|E''\rangle=\delta(E'-E'') \tag{4.2}$$

For the quantum systems with continuous spectra the dimensionless energy is $E$ (where we assume that $\beta_B=(k_B T)^{-1}$ is also dimensionless). So, we cannot talk about linear or quadratic spectrum.

By applying the *generalized discrete – continuous limit* $d\to c$ let's transcribe for the continuous case the results above obtained for the discrete case.

The generalized integral Mittag-Leffler function becomes

$$\lim_{d\to c} E^{\gamma,k}_{\beta,\alpha}(z)=\lim_{d\to c}\sum_{n=0}^\infty \frac{(\gamma)_{n,k}}{\Gamma_\alpha(\beta+\alpha n)}\frac{z^n}{n!}\quad\to\quad \tilde{E}^{\gamma,k}_{\beta,\alpha}(z)=\int_0^\infty dE\,\frac{(\tilde{\gamma})_{E,k}}{\tilde{\Gamma}_\alpha(\beta+\alpha E)}\frac{z^E}{\Gamma(E+1)} \tag{4.3}$$

$$(\tilde{\gamma})_{E,k}=k^E\left(\frac{\gamma}{k}\right)_E = k^E\,\frac{\Gamma\!\left(\dfrac{\gamma}{k}+E\right)}{\Gamma\!\left(\dfrac{\gamma}{k}\right)} \tag{4.4}$$

$$\tilde{\Gamma}_\alpha(\beta+\alpha E)=\alpha^E\,\frac{\Gamma\!\left(\dfrac{\beta}{\alpha}+E\right)}{\Gamma\!\left(\dfrac{\beta}{\alpha}\right)}\Gamma(\beta)=\alpha^E\left(\frac{\beta}{\alpha}\right)_E \Gamma(\beta)=\Gamma(\beta)(\tilde{\beta})_{E,\alpha} \tag{4.5}$$

The generalized integral Mittag-Leffler function is connected with integral generalized hypergeometric function:

$$\tilde{E}^{\gamma,k}_{\beta,\alpha}(z)=\frac{1}{\Gamma(\beta)}\,{}_1\tilde{F}_1\!\left(\frac{\gamma}{k};\frac{\beta}{\alpha};\frac{k}{\alpha}z\right) \tag{4.6}$$

defined as

$${}_1\tilde{F}_1\!\left(\frac{\gamma}{k};\frac{\beta}{\alpha};\frac{k}{\alpha}z\right)=\int_0^\infty dE\,\frac{\left(\dfrac{\gamma}{k}\right)_E \left(\dfrac{k}{\alpha}z\right)^E}{\left(\dfrac{\beta}{\alpha}\right)_E\,\Gamma(E+1)} \tag{4.7}$$

For the particular case, when $\alpha=\beta=\gamma=k=1$, we get successively the integral exponential or the *nu*-function []:

$$\tilde{E}^{1,1}_{1,1}(z)={}_1\tilde{F}_1(1;1;z)={}_0\tilde{F}_0(\,;\,;z)=\int_0^\infty dE\,\frac{z^E}{\Gamma(E+1)}=\tilde{e}(z)=\nu(z) \tag{4.8}$$

***Additional note 4:*** Do not confuse the *integral exponential function* $\tilde{e}(z)$ with the *exponential integral function* $\mathrm{Ei}(x)$, for real $x$, which is a special function on the complex plane, defined as

$$\mathrm{Ei}(x)=-\int_{-x}^\infty dt\,\frac{e^{-t}}{t}=\int_{-\infty}^x dt\,\frac{e^t}{t} \tag{4.9}$$

Therefore, we can define the *integral generalized Mittag-Leffler function* as generalized *nu*-function

$$\tilde{E}_{\beta,\alpha}^{\gamma,k}(z) = \int_0^\infty dE \frac{(\tilde{\gamma})_{E,k}}{\tilde{\Gamma}_\alpha(\beta+\alpha E)} \frac{z^E}{\Gamma(E+1)} \equiv v_{\beta,\alpha}^{\gamma,k}(z) \tag{4.10}$$

In particular, with this new notation, the usual *nu*-function is $v(z) \equiv v_{1,1}^{1,1}(z)$. In this sense, this is a new application of *nu*-function.

The eigenvalues for the continuous spectrum becomes:

$$\lim_{\substack{n \to E \\ \alpha,\beta,\gamma,k=1}} e(n) \equiv \lim_{\substack{n \to E \\ \alpha,\beta,\gamma,k=1}} n \frac{(\beta+\alpha(n-1))}{\gamma+k(n-1)} = E \tag{4.11}$$

Consequently, the action of the Hermitian operators $\mathcal{A}_-$ and $\mathcal{A}_+$ on the vectors $|E>$ results from the application of the *discrete – continuous limit* $d \to c$ on their discrete counterparts :

$$\mathcal{A}_-|E> = \sqrt{E}|E-1>, \quad \mathcal{A}_+|E> = \sqrt{E+1}|E+1>, \quad \mathcal{A}_+\mathcal{A}_-|E> = E|E> \tag{4.12}$$

as well as by successively applications of these operators on the vacuum states $|0>$ and $<0|$ :

$$\begin{Bmatrix}|E>\\<E|\end{Bmatrix} = \frac{1}{\Gamma(E+1)}\begin{Bmatrix}(\mathcal{A}_+)^E|0>\\<0|(\mathcal{A}_-)^E\end{Bmatrix} \tag{4.13}$$

In [Rom. Rep. Phys] *for the continuous spectrum we introduced a real dimensionless energy parameter* $\varepsilon > 0$, which is *not a quanta*, and can be interpreted as a suitable "jump unity" in the energy scale of continuous spectra. By equating to unity $\varepsilon = 1$, the system's energy may be written simply as $E = m$.

Consequently, the generalized discrete – continuous limit $d \to c$ can also be applied to the definition of Mittag-Leffler coherent states:

$$\lim_{d \to c}|z> = \lim_{d \to c} \frac{1}{\sqrt{E_{1,1}^{1,1}(|z|^2)}} \sum_{n=0}^\infty \sqrt{\frac{(1)_{n,1}}{\Gamma_1(1+n)}} \frac{z^n}{\sqrt{n!}}|n> = \frac{1}{\sqrt{\tilde{E}_{1,1}^{1,1}(|z|^2)}} \int_0^\infty dE \frac{z^E}{\Gamma(E+1)}|E> = |\tilde{z}> \tag{4.14}$$

respectively for the version that contains the operators:

$$|\tilde{z}> = \frac{1}{\sqrt{\tilde{E}_{1,1}^{1,1}(|z|^2)}} \tilde{E}_{1,1}^{1,1}(z\mathcal{A}_+)|0>, \quad |\tilde{z}> = \frac{1}{\sqrt{v(z)}} v(z\mathcal{A}_+)|0> \tag{4.15}$$

Using these definitions, we can rewrite in the integral version all the relations obtained in the previous section, containing the integral Mittag-Leffler CSs or *nu*-functions, deduced for the discrete spectrum.

The orthogonal propagator onto states $|\tilde{z}>$ is

$$|\tilde{z}><\tilde{z}| = \frac{1}{\tilde{E}_{1,1}^{1,1}(|z|^2)} \# \frac{\tilde{E}_{1,1}^{1,1}(\mathcal{A}_+ z)\tilde{E}_{1,1}^{1,1}(\mathcal{A}_- z^*)}{\tilde{E}_{1,1}^{1,1}(\mathcal{A}_+\mathcal{A}_-)} \# \tag{4.16}$$

To find the limit for $z \to 0$ of this orthonormal propagator, we will use the *inverse limit*, i.e. *the generalized continuous - discrete limits* $c \to d$ :

$$\lim_{\substack{z \to 0 \\ c \to d}} \tilde{E}_{1,1}^{1,1}(z) = \lim_{\substack{z \to 0 \\ E \to n}} \int_0^\infty dE \frac{z^E}{\Gamma(E+1)} = \lim_{z \to 0} \sum_{n=0}^\infty \frac{z^n}{n!} = 1 \tag{4.17}$$

We thus arrive at the correct expression of the vacuum propagator

$$|0><0| = \frac{1}{\#\tilde{E}_{1,1}^{1,1}(\mathcal{A}_+\mathcal{A}_-)\#} \tag{4.18}$$

The integration measure $d\tilde{\mu}(z) = \frac{d\varphi}{2\pi} d(|z|^2) \tilde{h}(|z|^2)$ is obtained beginning from the closure relation

$$\int d\tilde{\mu}(z) |\tilde{z}><\tilde{z}| = 1 \tag{4.19}$$

We have

$$1 = \frac{1}{\#\tilde{E}_{1,1}^{1,1}(\mathcal{A}_+\mathcal{A}_-)\#} \int_0^\infty d(|z|^2) \frac{\tilde{h}(|z|^2)}{\tilde{E}_{1,1}^{1,1}(|z|^2)} \int_0^{2\pi} \frac{d\varphi}{2\pi} \#\tilde{E}_{1,1}^{1,1}(\mathcal{A}_+ z)\tilde{E}_{1,1}^{1,1}(\mathcal{A}_- z^*)\# \tag{4.20}$$

$$\int_0^{2\pi} \frac{d\varphi}{2\pi} \#\tilde{E}_{1,1}^{1,1}(\mathcal{A}_+ z)\tilde{E}_{1,1}^{1,1}(\mathcal{A}_- z^*)\# = \int_0^\infty dE \frac{\#(\mathcal{A}_+\mathcal{A}_-)^E\#}{[\Gamma(E+1)]^2} (|z|^2)^E \tag{4.21}$$

so that the following integral must be

$$\int_0^\infty d(|z|^2) \frac{\tilde{h}(|z|^2)}{\tilde{E}_{1,1}^{1,1}(|z|^2)} (|z|^2)^E = \Gamma(E+1) \tag{4.22}$$

Making the substitution $E = s - 1$, this is the problem of Stieltjes moments [14], so the solution is

$$\frac{\tilde{h}(|z|^2)}{\tilde{E}_{1,1}^{1,1}(|z|^2)} = G_{0,1}^{1,0}(|z|^2|0) = e^{-|z|^2} \tag{4.23}$$

The final expression of the integration measure is, then

$$d\tilde{\mu}(z) = \frac{d\varphi}{2\pi} d(|z|^2) e^{-|z|^2} \tilde{E}_{1,1}^{1,1}(|z|^2) \tag{4.24}$$

It can also be obtained directly, if in Eq. (2.12) we put $\alpha = \beta = \gamma = k = 1$.

The corresponding density operator is, then

$$\tilde{\rho} = \frac{1}{\tilde{Z}(\beta_B)} \int_0^\infty dE\, e^{-\beta_B E} |E><E| \tag{4.25}$$

$$\tilde{\rho} = \frac{1}{\tilde{Z}(\beta_B)} \frac{1}{\#\tilde{E}_{1,1}^{1,1}(\mathcal{A}_+\mathcal{A}_-)\#} \#\tilde{E}_{1,1}^{1,1}(e^{-\beta_B}\mathcal{A}_+\mathcal{A}_-)\# \tag{4.26}$$

Taking into account equality

$$\tilde{E}_{1,1}^{1,1}(z) = \int_0^\infty dE \frac{z^E}{\Gamma(E+1)} = \nu(z) \tag{4.27}$$

all these formulas can be expressed by the *nu*-function.

The partition function $\tilde{Z}(\beta_B)$ is obtained from the normalization relation to unity of the density operator $\text{Tr}\tilde{\rho} = 1$:

$$\mathrm{Tr}\,\tilde{\rho} = \int_0^\infty dE'\,<E'|\tilde{\rho}|E'> = \frac{1}{\tilde{Z}(\beta_B)}\int_0^\infty dE'\int_0^\infty dE\,e^{-\beta_B E}\,<E'|E><E|E'> =$$
$$= \frac{1}{\tilde{Z}(\beta_B)}\int_0^\infty dE\,e^{-\beta_B E} = \frac{1}{\tilde{Z}(\beta_B)}\frac{1}{\beta_B} = 1 \quad (4.28)$$

Finally, we observed that the partition function is linear dependent on the temperature $\tilde{Z}(\beta_B) = k_B T$.

Consequently, the Husimi's distribution function becomes

$$\tilde{Q}(|z|^2) = <\tilde{z}|\tilde{\rho}|\tilde{z}> = \frac{1}{\tilde{Z}(\beta_B)}\int_0^\infty dE\,e^{-\beta_B E}\,<\tilde{z}|E><E|\tilde{z}> =$$
$$= \frac{1}{\tilde{Z}(\beta_B)}\frac{1}{\tilde{E}_{1,1}^{1,1}(|z|^2)}\tilde{E}_{1,1}^{1,1}(e^{-\beta_B}|z|^2) = \frac{1}{\tilde{Z}(\beta_B)}\frac{\nu(e^{-\beta_B}|z|^2)}{\nu(|z|^2)} \quad (4.29)$$

where you can see the application of $\mathcal{A}_+\mathcal{A}_- \to |z|^2$ correspondence.

Finally, the diagonal representation of the density operator is written

$$\tilde{\rho} = \frac{1}{\tilde{Z}(\beta_B)}\int d\tilde{\mu}(z)\,|\tilde{z}>\tilde{P}(|z|^2)<\tilde{z}| \quad (4.30)$$

After the appropriate substitutions, following the same step as in the case of deducing the expression of the integration measure, we obtain that the quasi-distribution function *P* has the following final expression:

$$\tilde{P}(|z|^2) = \frac{1}{\tilde{Z}(\beta_B)}\frac{G_{0,1}^{1,0}(e^{\beta_B}|z|^2\,|0)}{G_{0,1}^{1,0}(|z|^2\,|0)} = \frac{1}{\tilde{Z}(\beta_B)}\exp\left[-(1-e^{\beta_B})|z|^2\right] \quad (4.31)$$

***Additional note 5:*** Up to a constant (partition function), the $\tilde{P}(|z|^2)$ distributions coincide with the corresponding ones of the linear harmonic oscillator (HO-1D).

## 5. Concluding remarks

We defined and used a pair of Hermitian annihilation and creation operators which generate the generalized coherent states, defined in the Barut-Girardello manner. The choice of these operators was not accidental, the reason being that they generate such coherent states whose normalization function is just the four-parameter generalized Mittag-Leffler function. We examined the characteristic properties both for pure and mixed (thermal) coherent states. All calculations are made using the rules of the technique of diagonal ordering of operators which we introduced in a previous paper and which proved useful in performing the calculations. Finally, the integral counterpart of the Mittag-Leffler coherent states are examined which is connected with *nu*-function. In this way, the present paper becomes an example of a new application of a mathematical entity (Mittag-Leffler function) in quantum mechanics (coherent states formalism).